\documentclass[useAMS,usenatbib,usegraphicx]{mn2e}

\title{Visible and dark matter in M\,31 - I. Properties of stellar components}
\author[A. Tamm, E. Tempel and P. Tenjes]{A. Tamm$^{1}$\thanks{E-mail:
atamm@ut.ee;
elmo@aai.ee; peeter.tenjes@ut.ee}, E. Tempel$^{1,2}$ and P. Tenjes$^{1,2}$\\
$^{1}$Tartu Observatory, 61602 T\~oravere, Estonia\\
$^{2}$Institute of Theoretical Physics, Tartu University, T\"ahe 4,
51050 Tartu, Estonia}

\begin{document}

\date{Accepted 2008 Month 00, Received 2007 Month 00.}

\pagerange{\pageref{firstpage}--\pageref{lastpage}} \pubyear{2007}

\maketitle

\label{firstpage}

\begin{abstract}

We construct a structural model of the Andromeda Galaxy,
simultaneously corresponding to observed photometrical and
kinematical data and chemical abundances. In this paper we present
the observed surface brightness, colour and metallicity
distributions, and compare them to the model galaxy. In Paper~II \citep*{tempel:08} we present similar data for the kinematics, and derive the mass distribution of the galaxy.

To determine structural parameters of stellar components, we
have collected the observed {\it U\/}, {\it B\/}, {\it V\/}, {\it
R\/}, {\it I\/} and {\it L\/} luminosity distributions and have
calculated the related colour indices. By using far-infrared imaging
data of M\,31 and a thin dust disc assumption, we derive dust-free
surface brightness and colour distributions.

The model galaxy is constructed as a superposition of four
axially symmetric stellar components: a bulge, a disc, an inner halo
and an extended diffuse halo. A set of colour indices and
metallicity is ascribed to each component, consistent with chemical
evolution models of simple stellar populations, metallicity
distribution measurements and star formation history estimates. The
models of chemical evolution allow to estimate mass-to-light ratios
of the stellar components.

We find the total absorption corrected luminosity of  M\,31
to be $L_B = (3.3 \pm 0.7) \cdot 10^{10} \rmn{L_{\sun}}$,
corresponding to an absolute luminosity $M_B=-20.8\pm 0.2$\,mag. Of the
total luminosity, 41\,\% (0.57\,mag) is obscured from us by the dust
inside M\,31. The intrinsic colours of the stellar populations are
in agreement with an extended formation time of bulge and disc stars
6--10 gigayears ago, with an additional population of young stars
making up about 10\,\% of the disc luminosity and less than 1\,\% of
its mass. Halo colours are not well reproduced by chemical evolution
models; especially in the outer halo, the actual colours are
considerably redder than the modelled ones. The chemical evolution
model yields the following mass-to-light ratios: $M/L_B=$3.7--6.9
$\rmn{M_{\sun}/L_{\sun}}$ for the bulge, $M/L_B=$2.9--5.2
$\rmn{M_{\sun}/L_{\sun}}$ for the disc, $M/L_B=$2.7--5.7
$\rmn{M_{\sun}/L_{\sun}}$ for the inner halo and $M/L_B\approx 2$
$\rmn{M_{\sun}/L_{\sun}}$ for the outer diffuse halo. The total mass of the
visible matter is $M_{\rmn{vis}}=$(10--19)$\cdot 10^{10}
\rmn{M_{\sun}}$, giving the total intrinsic mass-to-light ratio of the
visible matter $M/L_B=$3.1--5.8\,$\rmn{M_{\sun}/L_{\sun}}$.

The use of the model parameters for a dynamical analysis and for
determining dark matter distribution is presented in Paper~II.

\end{abstract}


\section{Introduction}

Models of hierarchical merging in a cold dark matter (CDM) universe
describe rather well the observed large-scale structure of clusters,
superclusters and their network. They are also successful in
reproducing general properties of galaxies with different
morphological types \citep{abadi:03,bell:03b,nagamine:05,governato:07}.

On the other hand, the observed number of dwarf galaxies seems to be
too small in comparison with CDM models
\citep{klypin:99,moore:99,benson:02}. In the case of the Ursa Minor
dwarf spheroidal galaxy, the observed velocity dispersions are in
contradiction to the cuspy density profile of CDM haloes
\citep{kleyna:03}. But by far the most commonly referred
contradiction is related to central densities of dark matter halos.
Cosmological simulations generate DM haloes with central density
cusps of $\rho \sim r^{-1}$ or steeper
\citep*[e.g.][]{navarro:97,moore:99}. Observations of dwarf and low
surface brightness disc galaxies have usually shown that shallow
central density profiles fit the data better than cuspy profiles
\citep*{blaisouellette:01,borriello:01,deblok:02,salucci:03,weldrake:03,simon:05,
zackrisson:06,kassin:06,gentile:04,gentile:07,valenzuela:07}.

For a proper analysis of actual galaxies, it is necessary to know the
distribution of both visible and dark matter. Unfortunately, the
structure and mass distribution of stellar populations is not known
precisely enough. One of the most uncertain aspects here is the
mass-to-light ratio ($M/L$) of visible matter. In principle, chemical evolution
models allow to calculate the evolution of both luminosity and mass of a stellar
population, but they involve
several insufficiently constrained parameters (the Initial Mass Function,
starting time, duration and intensity of star formation in different subsystems,
initial metallicities etc). It is possible to decrease the degeneracy in $M/L$
determinations by using 3--5 different colour indices, but due
to measurement uncertainties, colour indices are often
controversial. Internal dust attenuation affects the surface
brightness distribution profiles and colour indices, causing
additional uncertainty in the distribution of visible matter.

A comparison of the results of galactic structure
modelling with additional and, moreover, independent observations
(metallicities of stellar populations, stellar rotation curve,
velocity dispersions along several slit-positions) allows us to
constrain the distributions of visible matter and thereafter DM.

In this work, we analyse the density distributions of visible and DM
components in a nearby luminous disc galaxy, the Andromeda galaxy
M\,31. The galaxy M\,31 was selected because (1)~photometrical and
kinematical (rotation, dispersions) data are known with sufficiently
high resolution in order to study the bulge region; (2)ˇvelocity
dispersions have been measured also outside the galactic apparent
major axis; (3)~direct measurements of stellar metallicities allow
to constrain $M/L$ of visible matter; (4)~independent
estimates of mass distribution on large scales are available
(globular clusters, satellites, stream, kinematics of the
Milky~Way\,+\,M\,31 system); (5)~in the case of the Milky~Way,
\citet{binney:01} derived that cuspy CDM haloes are inconsistent with the
observational data (but see a different study by \citet{battaglia:05}).

Composite models of galaxies (including M\,31), taking into account surface
photometry and the history of chemical evolution and kinematical data,
were first constructed by \citet{einasto:74c}. It was demonstrated
that these models allow to distinguish stellar populations in
galaxies and to calculate their main parameters. In addition
to this kind of general mass and luminosity distribution models, the
structure of M\,31 stellar populations have been studied by
\citet{hodder:95}, who adopted \citet{bahcall:84} model, and by
\citet{bellazzini:03}. A rather different kind of model on the basis
of phase-space distribution function was constructed by
\citet{widrow:05}.

In this paper, we construct a photometric model of M\,31 stellar
populations on the basis of surface brightness profiles in {\it
U\/}, {\it B\/}, {\it V\/}, {\it R\/}, {\it I\/}, {\it L\/} colours
\citep*[see also][]{tenjes:94a}. The derived photometrical model gives us
parameters of galactic components, colour indices among them.
Stellar metallicities are available from independent spectral
observations and colour-magnitude diagrams. We have used all these
data (component radii, luminosities, colour indices, metallicities)
as input parameters for chemical evolution models to calculate the
ages and $M/L$s of the components. The main output of
the present paper are density distribution parameters and
$M/L$s of visible galactic components.

In the companion paper \citep[Paper~II,][]{tempel:08}, we apply the
results of the present paper and construct a mass distribution model
of M\,31. Calculating the rotation velocities and velocity
dispersions of visible matter with the help of the dynamical model
\citep[see][]{tempel:06} we can estimate the amount of DM which must be
added to obtain an agreement with the measured rotation and
dispersion data.

We have applied the following general parameters in our
calculations: the inclination angle of M\,31 has been taken
77.5\degr \citep{walterbos:88,devaucouleurs:91}, the major axis position angle is
38.1\degr \citep{walterbos:87,ferguson:02} and the distance to M\,31 is
\hbox{785\,kpc} \citep{mcconnachie:05}, corresponding to the scale
\hbox{1\,arcmin = 228\,pc}. Absorption in the Milky~Way has been
taken according to \citet*{schlegel:98}.

\section{Observational data}

In this Section, we describe observational data used to construct a
photometrical model and to determine metallicities of galactic
components.

\subsection{Photometrical data} \label{obsphot}

By now surface photometry of M\,31 is available in {\it U\/}, {\it
B\/}, {\it V\/}, {\it R\/}, {\it I\/}, {\it J\/}, {\it H\/}, {\it
K\/}, {\it L\/} colours. References to earlier observations are
given in \citet{tenjes:94a}. In outer parts of the galaxy,
\citet{pritchet:94} and \citet{irwin:05} derived surface brightness
profiles along the minor axis. The {\it V\/}-profile measured on the
basis of star counts by \citet{pritchet:94} extends up to the surface
brightness 29.3\,$\rmn{mag\,arcsec^{-2}}$. They estimated also the axial
ratio of isophotes in outer regions. \citet{irwin:05} measured
surface brightness in {\it V\/} and {\it I\/} colours directly and
with the help of star counts and compiled profiles up to
31.3\,$\rmn{mag\,arcsec^{-2}}$ in {\it V\/} and up to
29.2\,$\rmn{mag\,arcsec^{-2}}$ in {\it
I\/}.

Recent observations with the Spitzer Space Telescope
\citep{barmby:06} have allowed to extend the photometry to the {\it
L\/}-passband. The {\it J\/}, {\it H\/}, {\it K\/}-profiles have a
rather limited spatial extent and resolution, and we decided not to
use them in here. The wavelength of the Galaxy Evolution
Explorer near-UV observations
\citep{thilker:05} is unfortunately too
much off the standard {\it U} for interpretation with chemical
evolution models and we have not used these data here.

\begin{figure}
\includegraphics[width=84mm]{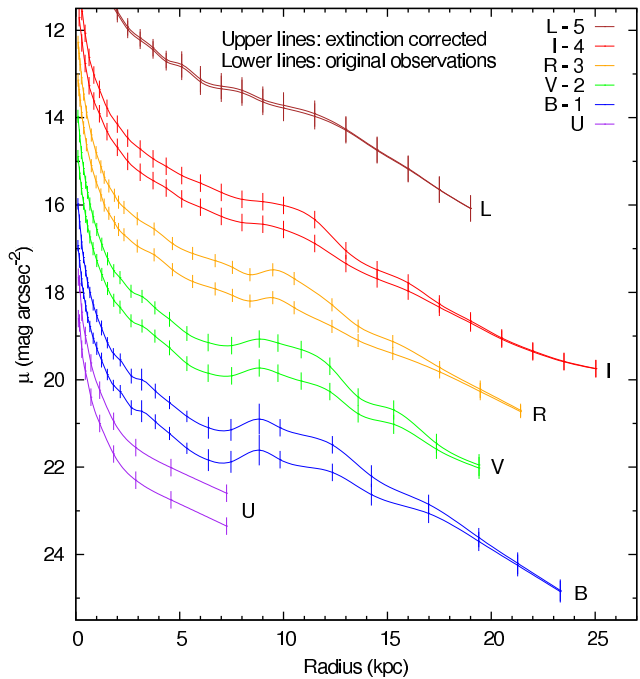}
\caption{M\,31 luminosity distribution along the major semiaxis in ,
{\it B\/}, {\it V\/}, {\it R\/}, {\it I\/}, {\it L\/} colours. For
each colour, the lower distribution corresponds to the measured data
points (Section~\ref{obsphot}), and the upper distribution presents our
surface brightness distribution corrected from the intrinsic
absorption (Section~\ref{dust}). Note that except for {\it U\/}, all other
distributions
have been shifted along the \emph{y}-axis as indicated in the
legend.}\label{neel}
\end{figure}

The composite surface brightness profiles in {\it U\/}, {\it B\/},
{\it V\/}, {\it R\/}, {\it I\/}, {\it L\/} colours along the major
and/or the minor axes were derived by averaging the results of
different authors. Different {\it R\/} and {\it I\/} colour system profiles were
transferred into the Cousins system, using the calibration by \citet{frei:94}. The
resultant observational profiles along the major axis together with the
estimated errors are presented in Fig.~\ref{neel}. All the surface brightness
profiles obtained in this way belong to the initial data set of our model
construction.

\subsection{Metallicity measurements}\label{metobs}

\begin{figure}
\includegraphics[width=84mm]{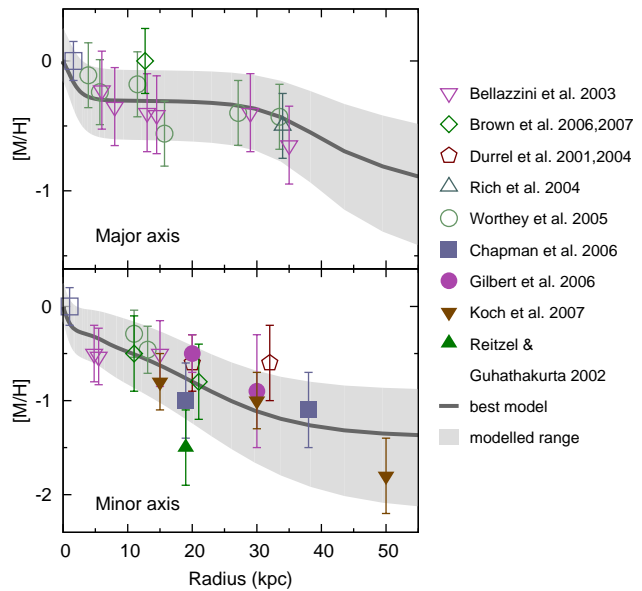}
\caption{ Metallicities of M\,31 along the major axis and the minor
axis. Open symbols represent photometric estimates, filled symbols
represent spectroscopic measurements. The solid line gives the
best-fit metallicity distribution of our model; the shaded region
shows the whole possible metallicity range, used to determine the
uncertainties of the modelled colours and mass-to-light ratios
(Section~\ref{chemmod}).}\label{met}
\end{figure}

To constrain the chemical evolution models, we have
collected from the literature most recent metallicity estimates at
various locations over the galaxy, extending from the central bulge
region to the outermost halo fields; the sources are listed in the
legend of Fig.~\ref{met}.  Within these works, the metallicities of
the dense bulge and disc regions have been estimated by fitting
isochrones to colour-magnitude diagrams of the red giant branch (RGB) stars
found in
the fields. In the outer low-density fields, isolating of M\,31
stars from the foreground Milky~Way dwarfs  becomes a crucial issue,
and spectroscopic measurements are necessary for higher reliability.
Furthermore, according to recent studies, the outer regions of
M\,31 contain a number of distinct stellar populations in the form
of stellar streams, which probably are remnants of former satellites
of M\,31 with their own distinct metallicity \citep{ibata:07}. Thus
metallicity estimates based on randomly selected outer fields may
introduce a significant scatter and even a bias.

In different papers, metallicity has been presented in both
iron-to-hydrogen ratio [Fe/H] as well as metal-to-hydrogen ratio
[M/H] (also called the ``global metallicity" scale). Photometric
properties of a star depend on the abundances of all heavy elements
rather than iron alone, thus we have converted all the measurements
into the [M/H] scale as a more informative one. Conversion from one
scale to another is not straightforward; for example, in the Milky~Way disc,
solar-metallicity stars have been shown to have
metal-to-iron ratio close to that of the Sun, while an enhancement
of the abundance of the $\alpha$-elements with respect to iron
occurs with a decreasing metallicity and thus with an increasing
radius \citep{edvardsson:93}; a similar enhancement has been found
for the Milky~Way halo stars \citep*[e.g.][]{zhao:90}. Following
\citet*{salaris:93} and \citet*{ferraro:06}, in the disc region we apply
a conversion $\rmn{[M/H]} = \rmn{[Fe/H]} + \log(0.638f_{\alpha}+0.362$), where
the
$\alpha$-element enhancement factor $f_{\alpha}=10^{-0.3\rmn{[Fe/H]}}$
within the range $0>\rmn{[Fe/H]}>-1$. In the halo fields, we apply
$\rmn{[M/H]} = \rmn{[Fe/H]} + 0.2$, assuming the $\alpha$-element enhancement by
a
factor of 2.

In Fig.~\ref{met}, open symbols represent photometric
metallicity estimates and filled symbols stand for spectroscopic
measurements. The data points in the upper and lower panel of
Fig.~\ref{met}  are given as projected along the major and minor
axis of the galaxy, respectively, accounting for the mean
ellipticity of the galaxy at the location of each metallicity
measurement. Due to differences in the geometry of the galactic
components, such a projection cannot be correct for fields located
too much off the axes onto which the projection is made, thus we
have not used data from disc fields deviating from the target axis
by angles more than about 20 degrees. The error bars of
metallicities present both the direct measurement uncertainties and
variations arising from the uncertainty of the distance to the
galaxy, reddening etc; see \citet*{bellazzini:03} and \citet{brown:06}
for an estimate of these effects. To consider all the uncertainties
of our model resulting from this rather vaguely known metallicity
distribution, we have used in our calculations the whole allowed
range of metallicities, presented with the grey area in
Fig.~\ref{met}.

\section{Surface brightness distribution model}

\subsection{Recovering dust-free luminosity distributions}\label{dust}

Before fitting any models to the observations it is
essential to  consider possible effects of dust extinction on the
observed data. Ignoring of dust effects may lead to substantially
wrong estimates of the structure, luminosity, colour and the mass of
the stellar components. Accounting for the effects of intrinsic dust
extinction, however, is a complicated task. Fortunately, the
Andromeda is an intensely studied galaxy and its dust distribution
has been revealed by surveys with \emph{IRAS}, \emph{ISO} and
\emph{Spitzer} satellites at a variety of far-infrared wavelengths.

Conversion of the infrared flux along each line-of-sight
into absorption is far from trivial, even if we assume the flux to
be proportional to the dust column density, and we know the
absorption properties of the dust. Also, the actual absorption along
each line-of-sight strongly depends on the geometry of the stellar
populations and the dust layer. The \emph{Spitzer} telescope
infrared view in \citet{gordon:06} reveals dust concentrated in
several spiral segments in the disc region of the galaxy, peaking at
about a dozen kiloparsecs from the centre.

Relying on reddening measured from colour histograms of
globular  clusters in M\,31 halo by \citet{barmby:00}, we have
assumed that the extinction law of M\,31 is similar to that of the
Galactic dust. To calculate the total absorption, we use optical
depth estimates from the literature. \citet{xu:96} constructed a
sophisticated model to estimate face-on optical depth of the M\,31
dust on the basis of \emph{UV}, optical and infrared luminosity, the
latter relying on the \emph{IRAS} measurements at 60\,$\umu$m and
100\,$\umu$m. They found the azimuthally averaged optical depth
$\tau_V$ to be in the range of 0.7--1.6 at 2--14\,kpc from the
centre. Later on, \citet{haas:98} added the \emph{ISO} satellite
measurements at 175\,$\umu$m. By transferring the infrared
emission into dust mass and the latter into optical depth, they reached
an optical depth value close to that of \citet{xu:96}. However, to
correct optical luminosity profiles, the uncertainties are large,
further increased by our limited knowledge about the 3-D
distribution and clumpiness of the dust clouds. While in the disc
region it is reasonable to consider dust being concentrated within a
rather thin dust disc as a result of disc kinematics, in the central
bulge regions the dust clouds are likely to move along more varied
orbits. A spherical rather than discy distribution of dust is
suggested also by the far-infrared images. Unfortunately, accounting
for the absorption by roughly spheroidally distributed dust is too
uncertain, thus we have avoided using the data within about 2
kiloparsecs from the centre along the minor axis.

The recovery of the major axis profiles turns out to be a
much simpler task. We assumed the dust disc to be thin compared to
the stellar disc, with its optical depth proportional to the
infrared flux, and calculated the absorption along each line-of-sight. Interestingly, the resultant absorption is rather insensitive
to the assumptions about the dust disc thickness and face-on optical
depth. The high inclination angle of the galaxy increases the
effective optical depth of the dust disc so much that nearly all the
optical signal coming from the other side of the dust layer become
absorbed, and naturally all the light from the stars between us and
the dust layer reaches us unaffected. Thus in the case of a thin
dust layer with respect to the stellar disc, along the major axis
absorption is effectively grey, being about 0.75\,mag almost
uniformly for  {\it U\/}, {\it B\/}, {\it V\/}, {\it R\/}, {\it I\/}
filters and for any optical depth within the range $\tau_V =
0.5-\infty$. The original and restored luminosity profiles are
presented in Fig.~\ref{neel}.

For comparison, \citet{driver:07} have derived integral absorption value
as a function of inclination angle statistically on the basis of over 10,000
galaxies. For the inclination angle of M\,31, the absorption is 0.75--0.85\,mag
for the bulge and 0.6--0.65\,mag for the disc in {\it B\/}-filter, which matches
well with our estimate of 0.75\,mag.

\subsection{Choice of components}\label{comp}

A satisfactory fit to the measured luminosity distribution of M\,31
can be achieved with just one or two S\'{e}rsic profiles if only
one-dimensional photometry, based on a single filter observations, is
used \citep[e.g.][]{irwin:05}. In the case of M\,31, much more
information is available: multi-colour photometry along both major
and minor axes, metallicity maps, kinematical data -- a
comprehensive model should attempt to fit all these observations. A
better fit can be achieved by increasing the number of components,
for example it has been revealed from the infrared maps of M\,31
that the dusty star-forming region actually consists of several
rings with offset centres and inclination angles \citep{gordon:06}.
However, the use of a large number of components would blur
our vision about the structure and the origin of the galaxy, thus
here we try to use a minimal number of components giving
simultaneously a satisfactory fit to all collected data.

While the classical galactic components of the bulge and the
disc are clearly visually detectable, and their existence and
structure can undoubtedly be confirmed by studying the morphology of
the galaxy and the kinematics of the embedded stars, classification
of the outer stars into distinct stellar components is much more
complicated. The outer regions are of extremely low density and the
foreground stars of the Milky~Way dominate in the observed fields.
Moreover, recent studies indicate the presence of a considerable
non-relaxed substructure around M\,31 in the form of tidal streams
and disrupted companions \citep{ibata:07}, the most prominent of
which is the Giant Steam.

Several observational studies of distant fields along the
minor  axis have found evidence for the existence of stellar
populations far beyond the bulge and the disc regions
\citep{irwin:05, ibata:07}. The outermost regions seem to be dominated
by a faint, diffuse, very extended metal poor halo, perhaps
embedding most of the whole Local Group \citep{ibata:07}. We refer to
this underlying population as the ``outer diffuse halo". Between the
radii from a few to about 25 kiloparsecs, recent studies suggest the
presence of a dynamically warm but slowly rotating stellar
population of stars with intermediate metallicity, which we
designate as the ``inner halo". Note that the same region has been
referred to as the ``extended disc" \citep{ibata:05,ibata:07}, the
``spheroid" \citep{brown:07} and the ``bulge" \citep{kalirai:06},
depending on the viewpoint. Earlier, when no information about the
underlying metal-poor population was available, this population was
simply referred to as the ``halo".

We have not attempted to include the Giant Stream in our
model.  The Giant Stream stars lie slightly off the minor axis of
the galaxy and thus do not contribute significantly to our
luminosity and kinematics data, while its expected total mass and
luminosity can safely be fitted within the uncertainties of our
results. Besides, our model can deal with axially symmetric stellar
populations only. For similar reasons, we also neglect the suggested
bar population and the elongated (double) nucleus of the galaxy.

For simplicity, in our model we ascribe a set of a chemical composition
and a formation time and subsequently a metallicity, $M/L$s and colour indices
homogeneously over each component. In
reality, the parameters listed above are all, to some extent,
radius-dependent. However, the usage of a superposition of several
components enables to compensate for such a simplification and
provides a fairly good fit to the actual observational data, as will
be shown below and in Paper~II.

\subsection{Model density distribution of the components}

To construct a sufficiently flexible and consistent model, allowing
to describe both the surface luminosity distribution of components
and their dynamics, we start from a spatial density distribution law
for individual components, which allows an easier fitting
simultaneously for light distribution and kinematics.

In such a model, the visible part of the galaxy is given as a
superposition of its individual stellar components. The
spatial density distribution of each visible component is
approximated by an inhomogeneous ellipsoid of rotational symmetry
with the constant axial ratio $q$ and the density distribution law
\begin{equation}
l (a)=l (0)\exp \left[ -\left( {a \over ka_0}\right)^{1/N} \right] ,
\label{eq1}
\end{equation}
where $l (0)=hL/(4\upi q a_0^3)$ is the central density and $L$ is
the component luminosity; $a= \sqrt{R^2+z^2/q^2}$, where $R$ and $z$
are two cylindrical coordinates; $a_0$ is the harmonic mean radius
which characterizes rather well the real extent of the component,
independently of the parameter $N$. The coefficients $h$ and $k$ are
normalizing parameters, depending on $N$, which allows the density
behaviour to vary with $a$. The definition of the normalizing
parameters $h$ and $k$ and their calculation is described in
Appendix~B of \citet{tenjes:94a}. The luminosity density distribution
(\ref{eq1}) proposed independently by \citet{einasto:69} is similar
to the \citet{sersic:68} law for surface densities. Differences
between Eq.~\ref{eq1}, the S\'{e}rsic law and their structure
parameters $N$ can be seen in \citet{tamm:05}.

The density distributions for the visible components were projected
along the line-of-sight, and their sum yields the surface brightness
distribution of the model
\begin{equation}
 L(A)= 2\sum_{i=1}^4\frac{q_i}{Q_i} \int\limits_A^\infty
\frac{l_i(a)a\,\rmn{d}a}{(a^2-A^2)^{1/2}} ,
\label{eq2}
\end{equation}

where $A$ is the major semiaxis of the equidensity ellipse of the
projected light distribution and $Q_i$ are their apparent axial
ratios $Q^2=\cos^2\Theta+q^2\sin^2\Theta$. The angle between the
plane of the galaxy and the plane of the sky is denoted by $\Theta$.
The summation index $i$ designates four visible components.

The model parameters $q$, $a_0$, $L$ and $N$ for the components
were determined by a subsequent least-squares approximation process.
First, we made a crude estimation of the population parameters. The
purpose of this step is to avoid obviously non-physical parameters
-- relation (\ref{eq2}) is non-linear and fitting a model to
the observations is not a straightforward procedure. Next, a
mathematically correct solution was found for each component.
Uncertainties  of the model (related to the coupling of parameters) were
estimated by using the partial second derivatives of the sum of
the least-square differences \citep{bevington:03}. Details of the
least-squares approximation and the general modelling procedure were
described by \citet{einasto:89a,tenjes:94a}; \citet*{tenjes:98}; Tempel (in
preparation).

The parameters of the model (the axial ratio $q$, the harmonic mean
radius $a_o$,  the structural parameters $N$, the dimensionless
normalizing constants $h$ and $k$, $UBVRIL\/$-luminosities) are given
in Table~\ref{model_param}.

\section{Stellar composition of the populations}

\subsection{Chemical evolution models}\label{chemmod}

The key objective of considering chemical evolution here is to have
an estimate for $M/L$s of the stellar components
independently of their kinematics. In principle, stellar
$M/L$s can be treated as free parameters in dynamical
galaxy models; masses of galactic components are then determined by
the best fit to the kinematical data. In reality, a high degree of
degeneracy exists in such a fitting of the components, e.g. the disc
mass can be raised if the dark halo mass is lowered. Thus the use of
$M/L$s, provided by chemical evolution models reduces
the uncertainties arising from such degeneracy.

In our model, galactic components are assumed to be homogeneous
stellar populations. For this reason, in order to describe chemical
evolution, we can use for each component a simple stellar population
(SSP) evolution model. In SSPs, stars are left to chemically evolve
according to the observationally determined evolutionary tracks
(colour-magnitude diagrams). The population is formed of
born-together stars of various masses; the relative amount of stars
of a given mass is determined by the given initial mass function
(IMF).

Note that unlike several other studies of both local and
high-redshift galaxies where a single SSP model has been adopted for
the whole galaxy, we use a separate SSP model for each component,
thus the galaxy as a whole is modelled as a superposition of several
SSPs which is closer to the actual situation.

A variety of SSP models have been developed elsewhere. We
have chosen to rely on the up-to-date and much-tested model
\emph{Starburst99}, designed at the Space Telescope Science
Institute initially for young stellar populations, but later adapted
also for modelling populations with ages up to the Hubble time
\citep{leitherer:99, vazquez:05}.

We constrain the metallicity of each population according to
the  measurements described in Section~\ref{metobs}. We further constrain the
models according to star formation history estimates from the
literature. Guided by \citet{olsen:06} and \citet{brown:06} we
suppose that the star formation, both in the bulge and in the disc,
started 10 gigayears ago and lasted with constant intensity for four
gigayears. In addition we pollute the disc with a younger generation
of stars having formed during the last four gigayears. According to the observed
coulours, we find the fraction of the younger population will be roughly 10\,\%
. Also referring to \citet{brown:06}, we set the
star formation of the inner halo having taken place 8--12 gigayears
ago and ascribe similar history to the outer halo.

SSP models are also strongly dependent on the choice of the input
IMF. While  being among the most important distribution functions in
astrophysics, the IMF, on the other hand is rather poorly known
despite the large amount of work devoted to it. On the basis of the
field star studies of the Milky~Way disc and star clusters,
relatively narrow constraints on the IMF have been imposed for stars
of at least the solar mass; much less is known about the low-mass
end of the function. Making the matters worse, there is no reason to
assume that all stellar populations in all galaxies descend from
identical IMFs; birth conditions may vary significantly with time
and environment, thus in the studies of other galaxies it is
important to consider the effect of possible deviations from the IMF
established for our own neighbourhood.  We use the 4-segment IMF by
\citet{kroupa:01} for population modelling, but also run the Monte
Carlo simulations to calculate variations resulting from the
provided IMF uncertainties.

\begin{figure}
\includegraphics[width=84mm]{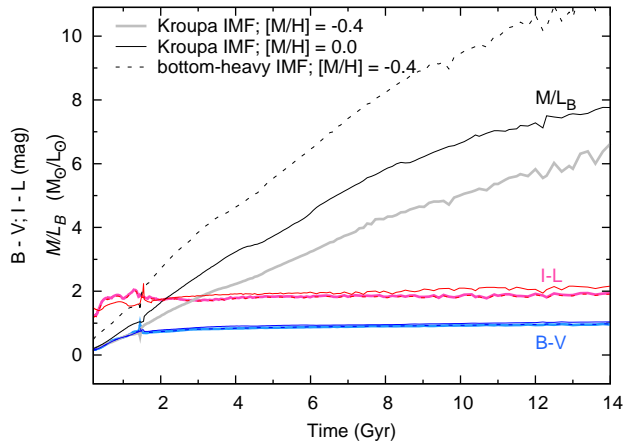}
\caption{Evolution of $B-V$ and $I-L$ colour indices and mass-to-light ratio $M/L_B$ at different metallicities and at different initial mass functions. Note that $M/L_B$ is very sensitive to variations of the initial mass function.}\label{chem}
\end{figure}
The effects of variations of age, metallicity and the IMF on the colours and mass-to-light
ratios of the modelled stellar populations are demonstrated in
Fig.~\ref{chem}. In this figure, the thick lines represent the evolution of a population with $\rmn{[M/H]} = -0.4$ (close to the average measured disc metallicity of M\,31) and with Kroupa IMF; the thin lines represent a population with the solar metallicity (rougly the upper limit of the disc metallicity). The population represented with the dashed lines has $\rmn{[M/H]} = -0.4$, but a ``bottom-heavy" IMF, giving rise to a higher number of low-mass stars and less high-mass stars (within the estimated uncertainties of the IMF). Fig.~\ref{chem} shows that uncertainties of the IMF have more effect on $M/L$ than uncertainties of metallicity determination, and that colour indices are rather independent of both of them. The resultant uncertainties of the colours and mass-to-light ratios of the actual M\,31 populations are demonstrated in Fig.~\ref{bulgedisc} and Fig.~\ref{haloes} with the
thickness of the model lines.

\subsection{Model fitting}

In a number of studies, $M/L$ or its distribution has been
determined on the basis of just a single colour estimate. When the
actual uncertainties of colour measurements are considered, the
derived $M/L$ cannot be very reliable. A single colour value can be
fitted with a large variety of initial metallicity, age, IMF, etc.
combinations. On the other hand, the use of more than one colour
measurement for determining the $M/L$ often indicates that
observations are actually controversial with the predicted colours
and the $M/L$ of the best-fitting model actually depends on which
colour is used for fitting. Thus we have used the colour information of all the
available observations from {\it U} to {\it L} to decrease such uncertainties,
giving in total five independent colour-indices.

The best-fitting model of the galaxy was found in an iterative
process. First a model density distribution, formed as a sum of the
four components described in Section~\ref{comp} was
least-squares-fitted to the observed luminosity distributions. This
was done simultaneously for all the available colours, allowing us
to determine the first-order approximation of model parameters and
providing us with colour indices of each component. These colour
indices were compared to those of the chemical evolution model for
the corresponding metallicity of the component, thus determining the
first-order $M/L$s for each colour. In the next
iteration, these $M/L$s were fed to the luminosity
distribution model, fixing the ratios of luminosities in different
filters for each component and resulting in slight corrections for
the component parameters and also the model colour-indices, which
were again compared to the chemical evolution models. After four or
five iterations, the changes of the parameters became negligible.

\begin{figure}
\includegraphics[width=84mm]{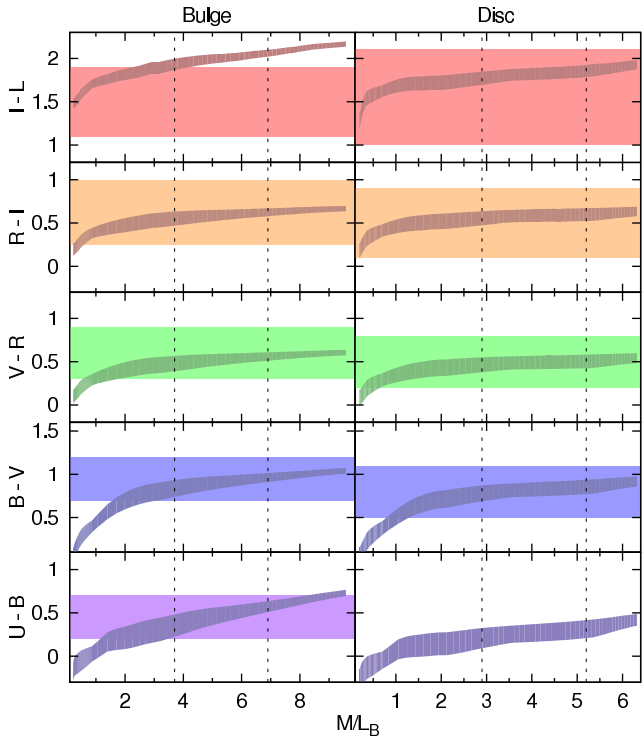}
\caption{Integrated colours of M\,31 bulge and disc components as a
function of mass-to-light ratio $M/L_B$. Filled areas -- constraints
from observations; darker areas -- predictions from chemical
evolution models  with uncertainties; dashed vertical lines -- limits
set by population age estimates.}\label{bulgedisc}
\end{figure}

\begin{figure}
\includegraphics[width=84mm]{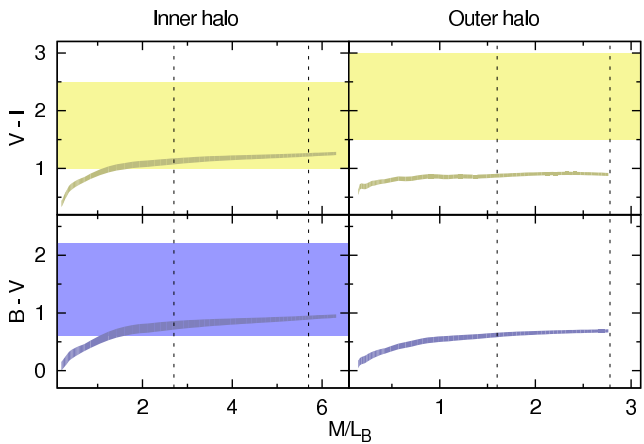}
\caption{The same as Fig.~\ref{bulgedisc} but for the halo
populations.  Note that especially in the outer halo, the modelled
colour stays considerably bluer than the observed one.}\label{haloes}
\end{figure}

Final results of \emph{Starburst99} chemical evolution model fitting
are presented in Fig.~\ref{bulgedisc} and Fig.~\ref{haloes}
with  darker lines. The wide regions of a constant colour represent
the observed colours of each population and the vertical dashed
lines in these figures denote upper and lower limits of the
acceptable $M/L$ values within the age limits described in
Section~\ref{chemmod}. In general, the colours predicted by the chemical
evolution model are in good concordance with the observed ones for
all the components with the exception of the diffuse outer halo.
Fig.~\ref{heledus} also shows that in general, the model colours
are not contradicting with the observed ones, apart from some
measurements in the outer regions along the minor axis. In the case
of the diffuse halo, the ``photometry" is actually based on the
counts of RGB stars and is probably not very accurate.

\section{Results and discussion}\label{Results}

\begin{figure*}
\includegraphics[width=150mm]{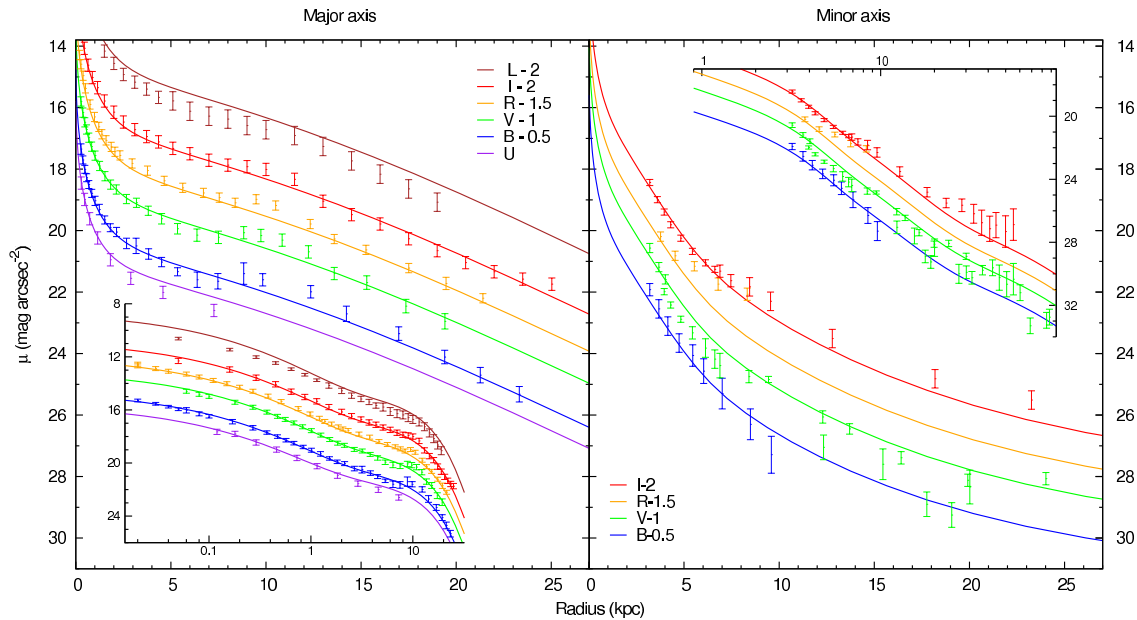}
\caption{Luminosity distribution of M\,31: left-hand panel --
$UBVRIL$ (from bottom to top) luminosity along the major axis;
right-hand panel -- $BVRI$ (from bottom to top) along the minor axis.
Error bars -- observations; solid lines -- model.}\label{heledus}
\end{figure*}

The absorption-corrected composite surface brightness profiles in {\it U\/},
{\it B\/},
{\it V\/}, {\it R\/}, {\it I\/}, {\it L\/} colours along the major
and the minor axes were derived by averaging the results of
different authors and are presented in Fig.~\ref{heledus} with errorbars. All
the surface
brightness profiles obtained in this way belong to the initial data
set of our model construction. With solid lines in Fig.~\ref{heledus},
we
show how the model fits to the restored luminosity distributions.
The four-component photometrical best-fitting model fits all
photometric profiles with a mean deviation $\langle \mu^{\rmn{obs}} -
\mu^{\rmn{model}}\rangle = 0.25$\,mag. The parameters of this model
(the harmonic mean radius $a_0$, the axial ratio $q$, the structural
parameters $N$, the dimensionless normalizing constants $h$ and $k$,
$UBVRIL\/$-luminosities) are given in Table~\ref{model_param}, together with
uncertainty estimates.

\begin{table*}
\caption[ ]{Calculated parameters of the photometrical model.}
\label{model_param}
\begin{flushleft}
\begin{tabular}{lllllll}
\noalign{\smallskip} \hline
Population        & $a_0$ (kpc)& $q$  & $N$ & $k$   & $h$   &   \\
\noalign{\smallskip} \hline
Bulge     &0.64$\pm0.03$& 0.6$\pm0.03$&4.2$\pm0.3$ & $6.498\cdot10^{-5}$& 4943 &
  \\
Disc      & 9.3$\pm0.3$ &0.05$\pm0.03$&0.7$\pm0.07$& 0.8479             & 2.240
&  \\
Inner halo& 4.0$\pm0.4$ & 0.5$\pm0.04$&2.7$\pm0.3$ & $7.229\cdot10^{-3}$& 159.8
&  \\
Outer halo& 40.         & 0.9         &2.0         & 0.05               & 33.33
&  \\
\noalign{\smallskip} \hline
Population       &  $L_U$ ($10^{10}\rmn{L_{\sun}}$) & $L_B$
($10^{10}\rmn{L_{\sun}}$) & $L_V$ ($10^{10}\rmn{L_{\sun}}$) & $L_R$
($10^{10}\rmn{L_{\sun}}$) &  $L_I$ ($10^{10}\rmn{L_{\sun}}$)& $L_L$
($10^{10}\rmn{L_{\sun}}$) \\
\noalign{\smallskip} \hline
Bulge     & 0.78$\pm0.1$& 1.09$\pm0.1$ & 1.45$\pm0.2$ & 1.70$\pm0.2$ &
2.28$\pm0.4$ & 6.62$\pm2.4$   \\
Disc      & 1.67$\pm0.5$& 1.68$\pm0.2$ & 2.02$\pm0.4$ & 2.31$\pm0.4$ &
3.11$\pm0.6$ & (7.88)   \\
Inner halo& (0.44)      & 0.50$\pm0.2$ & 0.61$\pm0.1$ & 0.69$\pm0.1$ &
0.88$\pm0.1$ & (1.86)  \\
Outer halo& (0.05)      & 0.05 & 0.05 & 0.06 & 0.06 & (0.10)  \\
\noalign{\smallskip} \hline
\end{tabular}
\end{flushleft}
\begin{list}{}{}
\item[] Notes: Luminosities are corrected for intrinsic absorption and
absorption in the Milky~Way. Luminosities given in parentheses are not based on
observed surface brightness profile; they are extrapolated according to the
chemical evolution models. Errors for the outer halo are undetermined.
\end{list}
\end{table*}

\begin{figure}
\includegraphics[width=84mm]{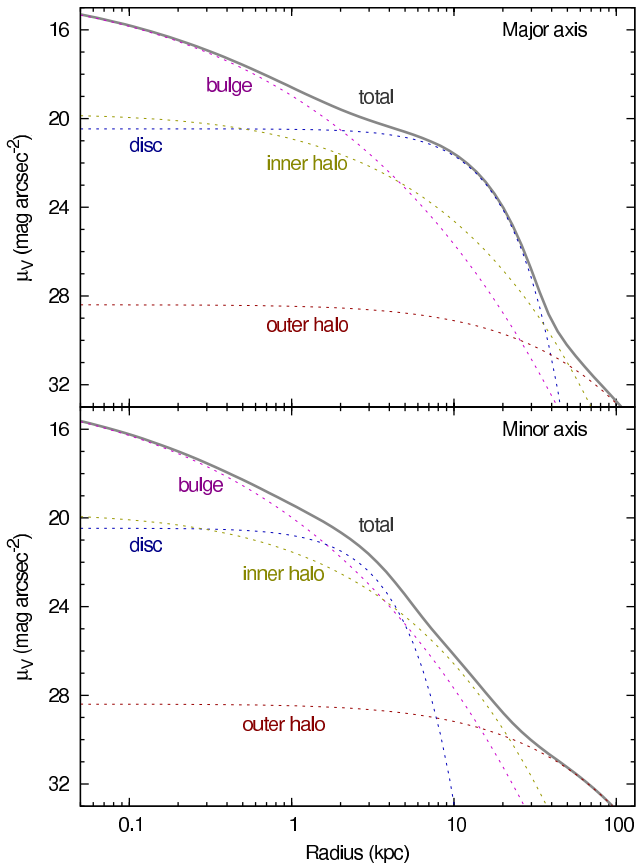}
\caption{Model surface brightness distribution in $V\/$-filter along the
major axis and minor axis. The
dashed lines indicate the contribution of each stellar population.
}\label{compfig}
\end{figure}

\begin{figure}
\includegraphics[width=84mm]{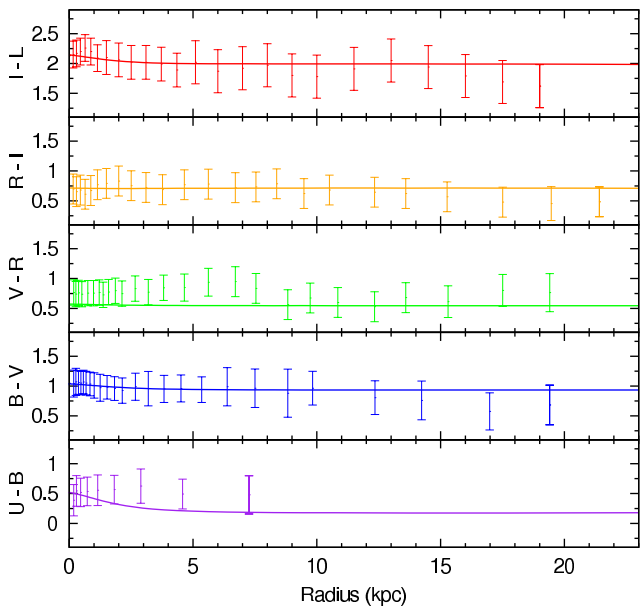}
\caption{Colour distribution in M\,31 along the major axis. From
bottom to top: $U-B$, $B-V$, $V-R$, $R-I$ and $I-L$. Error bars --
observations; solid lines -- model.}\label{colours}
\end{figure}

Fig.~\ref{heledus} shows the modelled luminosity distributions
along both the major and the minor axis. The contribution of the
individual components to the luminosity in $V\/$-filter is presented in
Fig.~\ref{compfig}. In Fig.~\ref{colours} the modelled colour
distribution is plotted.

The total luminosity of M\,31 corrected for the intrinsic absorption
is $L_B = (3.3 \pm 0.7) \cdot 10^{10} \rmn{L_{\sun}}$, corresponding to an absolute luminosity of $M_B=-20.8 \pm
0.2$\,mag. About half of it (51\,\%) is radiated by the disc, one third
(34\,\%) by the bulge, about 15\,\% by the inner halo and about one per
cent can be ascribed to the outer halo. The dust disc absorbs as
much as 41\,\% (or 0.57\,mag) of the total \emph{B}-flux. Of the total
apparent \emph{B}-luminosity, the disc, the bulge, the inner halo
and the outer halo give 51\,\%, 30\,\%, 17\,\% and 2\,\%, respectively. The
total intrinsic colour indices are $(B-V)=0.9 \pm 0.2,$ $(V-R)=0.6
\pm 0.2,$ and $(R-I)=0.7 \pm 0.3$. Colour indices predicted partly
by the chemical evolution model are $(U-B)=0.4$ and $(I-L)=1.9$.
Resulting from the chemical evolution model mass-to-light ratios
$M/L_B$ of the visible components together with the corresponding
ages and metallicities are given in Table~\ref{ml}.

The total mass of the visible matter $M_{\rmn{vis}}=$ (10--19)$\cdot
10^{10} \rmn{M_{\sun}}$, making the mean mass-to-light ratio of
the visible matter $M/L_B=$ 3.1--5.8\,$\rmn{M_{\sun}/L_{\sun}}$.

The main difference from a conventional bulge + disc + halo
model  is a separation of the inner, slowly rotating, moderately
enriched inner halo and the very metal-poor, diffuse and extremely
extended outer halo. This structure follows from the recent surface
brightness and kinematics measurements of the outer parts of M\,31,
which have revealed a slowly rotating component. The inner
halo in the present model is actually similar to the population
identified as the whole halo in previous stellar population
studies of M\,31 \citep{vandenbergh:91,tenjes:94a,geehan:06}. The outer diffuse
halo gives only marginal contribution to the overall luminosity of
the galaxy.

In their analysis of star counts, \citet{ibata:05, ibata:07} derived
the scale-length of the exponential ``extended disc'' along the
major axis to be $r_{\rmn{d}} =$ 6.6\,kpc. To match this value with the
minor axis scale length $r_{\rmn{d}} =$ 3.22\,kpc, they propose that the
extended disc component might have an inclination angle of
$60.8\degr$ instead of $i = 77\degr$ adopted for the rest of the
galaxy.  Our model suggests  the ``extended disc" (the ``inner halo" in
our model) to be a
thick component with an axial ratio of about $q = 0.5$,
which is close to the value $q \approx 0.55$ estimated for the
outer parts of M\,31 by \citet{pritchet:94}. Combined with
the measured ``warm" kinematics rather than ``cold" rotation
\citep{gentile:07}, we consider an ``inner halo" to be a more
descriptive name for this population.

The total stellar mass of M\,31, as estimated on the basis
of the $3.6\,\umu$m infrared luminosity by \citet{barmby:06},
is $11 \cdot 10^{10} \rmn{M_{\sun}}$. This lies within our estimated
range  (10--19)$\cdot 10^{10} \rmn{M_{\sun}}$. A more detailed mass
distribution model of M\,31 was constructed by \citet{geehan:06}, who
derived $M = 3.2\cdot 10^{10} \rmn{M_{\sun}}$ for the mass of the
bulge, and $7.2\cdot 10^{10} \rmn{M_{\sun}}$ for the mass of the
disc. They also derived $M/L_R =$ $3.9\, \rmn{M_{\sun}/L_{\sun}}$
for the bulge, and $M/L_R =$ $3.3,\ \rmn{M_{\sun}/L_{\sun}}$ for the
disc (not corrected for extinction). These values also lie within the range of
our model values: bulge mass (4--7.5)$\cdot 10^{10} \rmn{M_{\sun}}$, disc mass
(4.8--8.7)$\cdot 10^{10} \rmn{M_{\sun}}$ and mass-to-light ratios, not corrected
for extinction  $M/L_R=$ (2.4--4.4)\,$\rmn{M_{\sun}/L_{\sun}}$ for the
bulge and $M/L_R=$ (1.9--3.4)\,$\rmn{M_{\sun}/L_{\sun}}$ for the disc.

Fig.~\ref{bulgedisc} demonstrates that the colours predicted by
chemical evolution models match well with the observed
colours within age limits estimated in the literature. Thus we can
conclude that chemical evolution models and age estimates on the
basis of colour-magnitude diagrams are generally in very good
agreement with the actual integrated colours of M\,31. However, this
is not the case for the halo regions.  Fig.~\ref{haloes} shows
that especially for the outer halo the observed colours are
remarkably red compared to the modelled ones; even after a Hubble
time of evolution, the colours of the modelled population remain
significantly bluer. What could be the source of this discrepancy?
Quite clearly, the presence of dust in quantities required for
reddening at the level of $E(B-V)>0.5$\,mag is very unlikely in the
halo region. On the other hand, this would not be the first case of
finding evidence for the presence of extremely red haloes: the
stacked image of more than a thousand Sloan Digital Sky Survey disc galaxies
shows
significant {\it I\/}-colour excess \citep*{zibetti:04}, while \citet{zackrisson:06}
have found extremely red haloes around blue compact galaxies. In the
latter work, a bottom-heavy IMF was given as the most likely cause
for the effect.

\begin{table}
\caption[ ]{Properties of stellar populations, set by observations
and chemical evolution models.} \label{table_chem}
\begin{flushleft}
\begin{tabular}{llll}
\noalign{\smallskip} \hline
Population & Age (Gyr) & [M/H] & $M/L_B$ \\
\noalign{\smallskip} \hline
Bulge         &  6--10                  &  ~0.0  & 3.7--6.9     \\
Disc           &  6--10\,$^{\sharp}$    & -0.3  & 2.9--5.2     \\
Inner halo   &  8--12                  & -0.7  & 2.7--5.7     \\
Outer halo  &  8--12                  & -1.5  & 1.6--2.8\,$^{\flat}$  \\

\noalign{\smallskip} \hline
\end{tabular}
\label{ml}
\end{flushleft}
\begin{list}{}{}

\item[] $^{\sharp}$ Our model disc also contains 10\,\% of younger stars.
\item[] $^{\flat}$ $M/L_B$ of the outer halo is based on the chemical evolution
model
only.
\end{list}
\end{table}

The general ages of the bulge and disc stars suggest a similar
formation  history for these components, starting a few gigayears
after the Big Bang and continuing over an extended period. The small
additional younger population of stars, needed to achieve the
somewhat bluer colours of the disc, are in agreement with a
relatively recent merger event, perhaps with a structure which by
now has been disrupted to become the Giant Stream.

How unique is the developed model? To some extent, and
partly  because of the limitations set by the assumptions of
homogeneous populations with axial symmetry and avoidance of
ring-like structures, the components of the model are degenerate.
With suitably chosen structural parameters, the spheroidal bulge and
inner halo populations can compensate for each other. In fact, we
can think of the bulge and the inner halo as a single component with
a declining metallicity gradient towards outer radii. Similarly, the
actual structure of the two halo populations is by no means fixed by
our model, although there seems to be rather little in common in
their nature and origin. The error estimates for our final model are
given in Table~\ref{table_chem}. These estimates are based on the
partial second derivatives of the sum of the least-square differences
(the second derivatives of the goodness-of-fit parameter, see
\citet{bevington:03}, eq. 8.11). These errors describe how accurately
each model parameter is determined on the basis of the used
observational data. For the outer halo, where the observed
photometry was very uncertain, these estimates were quite large
(over 100\,\%) and we do not present them in Table~\ref{table_chem}.
The presented errors describe the goodness of our final model, and they do not
fully take into account the coupling of parameters.

We have attempted to estimate the effect of all observational and
methodological errors and uncertainties on our results. The dominant
source for uncertainties comes from the estimations of the
mass-to-light ratio, given by chemical evolution models. We have
tried to combine all the available information on the photometry and
metallicity of M\,31 and to use up-to-date evolution models of
stellar populations to minimize these uncertainties. But at a
critical glance, we are still left with masses and mass-to-light
rations uncertain by a factor of 10--30 per cent. These ranges can
be tightened with the inclusion of kinematic information. This is
done in Paper~II.

\section*{Acknowledgments}
We thank Dr.~J.~Pelt for a useful discussion about how to improve the fitting process in our
photometrical model and how to estimate the errors for the
parameters of the model. We acknowledge the financial support from
the Estonian Science Foundation {grants 6106, 7115, 7146} and project
SF0060067s08.
All the figures were made with the GNUPLOT plotting
utility.


\label{lastpage}
\end{document}